\documentclass[twocolumn,prl]{revtex4}
\usepackage{graphicx}% Include figure files
\usepackage{dcolumn}% Align table columns on decimal point
\usepackage{bm}% bold math

\newcommand{\beq}{\begin{equation}}
\newcommand{\eeq}{\end{equation}}
\newcommand{\beqa}{\begin{eqnarray}}
\newcommand{\eeqa}{\end{eqnarray}}

\def\l({\left(}
\def\r){\right)}
\def\lf[{\left[}
\def\ri]{\right]}

\begin{document}

\title{On the mechanism of prompt emission of gamma-ray bursts.}

\author{Andrey Neronov}

\affiliation{Laboratoire de physique des particules et cosmologie, EPFL, \\
BSP-Dorigny 1015, Lausanne, Switzerland}

\begin{abstract}
We propose a model in which prompt $\gamma$ emission of gamma-ray
bursts is the synchrotron radiation of electron-positron plasma in the ordered 
magnetic field in the direct vicinity of horizon of a young black hole 
formed in the core collapse of a massive star. This mechanism can naturally 
explain high degree of polarization of the $\gamma$-ray flux and hard low 
energy photon spectral index. Interaction 
of $\gamma$-quanta with ambient matter
provides a mechanism of formation of relativistic ejecta  
which are responsible for the $\gamma$-ray burst afterglows.     
\end{abstract}

\maketitle

%\pacs{PACS: 98.62.Nx, 98.54.Gr, 98.54.Cm, 96.40.-z}

Recently established connection between gamma-ray bursts (GRB) and Type Ic 
supernovae \cite{grb-sn,grb-sn1} supports the assumption that long 
GRBs can  be produced in the core collapse of  massive 
stars \cite{paczynski,woosley}. The collapse 
 of high mass ($M_\star>40 M_\odot$) stars can
result in direct production of black holes of masses $M_\bullet\sim
10 M_\odot$ \cite{hegeretal}. 
Subsequent accretion of the stellar material onto the newly born black hole
can serve as a ``central engine'' of GRBs (see e.g. 
\cite{paczynski}).  

In most of the GRB models (which can be called ``fireball-type'' models, 
see \cite{grb-review} for a recent review) 
the nature of the central engine of a GRB 
remains unclear because  prompt $\gamma$ emission 
is produced at a large distance  from the center. This is unavoidable 
if one supposes that  total energy of a GRB is deposited 
into kinetic energy of particles 
in a small volume of the size about the size of 
central engine. In this case particle density is so high that the 
optical depth for the $\gamma$-ray photons is $\tau\gg 1$. 
The spectrum and temporal 
characteristics of the prompt $\gamma$ emission are not related to the 
properties of the central engine, but are, in fact, 
determined by the properties of the shocks at a distance 
$R_{fin}> 10^{14}$ cm \cite{grb-review} at which matter becomes transparent 
to  $\gamma$ rays. 

The fireball-type models of 
GRBs have certain problems
in explaining characteristics of the prompt $\gamma$-ray flux, such as 
the very steep low-energy spectral index $\alpha$ of BATSE GRBs \cite{preece} 
or the high degree of polarization  of prompt $\gamma$ emission detected by
RHESSI \cite{rhessi} (see \cite{granot,waxman} for discussion of this
subject). Both problems arise because the emission
region is not compact and it is difficult to expect that the 
$\gamma$ ray synchrotron emission can be self-absorbed or the magnetic field
can be ordered on the distance scale $\sim R_{fin}$. 

The fact that  properties of $\gamma$-ray emission are unrelated to 
the physics of the central engine holds only for the models in which GRB energy
is stored at some stage 
as kinetic energy of relativistic particles (in this respect 
models which involve pulsars \cite{usov}, or electromagnetic 
outflows from a black hole \cite{lyutikov} 
are similar to the fireball model). However,
if the central engine of GRB 
is a black hole, the energy can be stored in the form of 
the black hole rotational energy \cite{paczynski}. The problem of 
``compactness'' can be overcome if one considers a mechanism of conversion
of this rotation energy directly into $\gamma$-ray flux, without intermediate
``particle dominated'' phase. In this case 
the prompt emission of a GRB can originate from
a very compact region of the size about the gravitational radius of a black
hole (one still has to assume sufficient beaming of the $\gamma$ ray flux 
so that the pair production in $\gamma\gamma$ 
interactions is suppressed). 
In such scenario the total energy, duration and variability of GRB would 
be directly related to the physical parameters of the central engine.

 In what follows 
we propose a model in which prompt emission of a GRB is produced in the 
vicinity of horizon of a young stellar mass black hole. We show that 
combined effect of pair production by $\gamma$ quanta in strong magnetic field 
and absorption of particles by the black hole horizon can lead to formation 
of a relatively low density layer of $e^+e^-$ plasma  close to the  
horizon. Electrons and positrons accelerated in the vicinity
of horizon by rotation-induced electric field 
suffer from severe energy losses so that all the work done by external field
is immediately radiated in the form of MeV $\gamma$ rays. The 
$\gamma$ rays are  emitted in a highly anisotropic way: 
in a form of two oppositely directed 
beams with the opening angles $\theta\sim 1/\Gamma_e$ where $\Gamma_e\sim 
10\div 100$ is the typical gamma-factor of electrons. Subsequent 
interaction of the $\gamma$-ray beams with  background matter 
can give rise to a rapid deposition of a fraction of the beam energy 
into the surrounding matter and production of relativistic ejecta 
much like in the fireball models. Thus, the model under consideration retains 
the advantages of the ``standard'' model of GRB afterglows but 
relates the prompt $\gamma$ emission to the mechanism of  production of 
relativistic outflows close to the black hole.  
   
Our set up is close to the one considered in \cite{paczynski,woosley}: the core
of a massive, rapidly rotating star collapses to form a  
$M_\bullet\sim 10M_\odot$ black hole. The outer parts of the core can form 
an accretion disk/torus which can support strong magnetic field 
$B_\bullet\ge 10^{12-13}$ G.  An estimate of the strength 
of magnetic  field can be obtained from the
``equipartition argument'' (assumption that the energy densities of 
the magnetic field and radiation are of the same order). If the typical 
luminosity of the central engine is $L_\bullet\sim 10^{48}$ erg/s and 
the size of the  emission region is  
$R\sim R_\bullet$, where $R_\bullet=2GM_\bullet=
3\times 10^6(M_\bullet/M_\odot)$ cm, the equipartition 
magnetic field is 
\beq
\label{equi}
B_\bullet\sim 10^{12}\lf[\frac{L_\bullet}{10^{48}\mbox{erg/s}}\ri]^{1/2}
\lf[\frac{10M_\odot}{M_\bullet}\ri]\mbox{ G}
\eeq
(much stronger magnetic field was supposed in 
\cite{paczynski} because the assumed mechanism of extraction of 
rotational energy of the black hole  was different from the one considered 
below).

If the magnetic field is produced by the accretion disk/torus, the 
typical scale 
of spatial variations of the field is about the size of the disk  which is 
larger than the size of the black hole. The magnetic field
is ordered at the length scale $R\sim R_\bullet$ \cite{membrane}. 
Synchrotron radiation from electrons and/or positrons in this 
magnetic field 
can provide reasonable explanation for the observed polarization of the prompt 
$\gamma$-ray emission. 

The energy of electrons which emit in  MeV 
band in the above magnetic field is 
\beq
\label{ee}
E_e=7.3\times 10^6\lf[\frac{\epsilon_\gamma}{1\mbox{ MeV}}\ri]^{1/2}
\lf[\frac{10^{12}\mbox{ G}}{B_\bullet}\ri]^{1/2}\mbox{ eV.}
\eeq 
The propagation distance of such electrons is just about 
$\lambda_{sy}=4\times 10^{-7}\lf[10^{12}\mbox{ G}/B_\bullet\ri]^{3/2}
\lf[1\mbox{ 
MeV}/\epsilon_\gamma\ri]^{1/2}$cm $\ll R_\bullet$.
This means that electrons should be continuously accelerated near the black 
hole horizon. 

Electric field which accelerates electrons can be generated by various 
mechanisms by the accretion disk and by the black hole itself. 
In fact, rotational 
drag of external magnetic field near horizon leads to the generation of 
electric 
field whose strength is $E_\bullet\sim (a/M_\bullet)B_\bullet$ 
($a\le M_\bullet$ 
is the rotation moment per unit mass) 
\cite{membrane}. Electron 
acceleration rate in the electric field whose strength is comparable to 
$B_\bullet$ can be estimated as $dE_e/dt=qeB_\bullet$ where 
$q\le 1$ is the 
parameter which characterizes effectiveness of the acceleration 
process. Maximal energy of electrons is determined by the balance between
acceleration and energy loss rates. If electric field is not
aligned with  magnetic field, the energy loss rate 
coincides by order of magnitude with synchrotron loss rate \cite{landau}. 
In this case the  maximal energy of electrons is
\beq
\label{emax}
E_{max}=\left[\frac{3qm_e^4}{2e^3B_\bullet}\right]^{1/2}\sim 3\times 10^{7} 
q^{1/2}\left[\frac{10^{12}\mbox{ G}}{B_\bullet}\right]^{1/2}
\mbox{ eV.}
\eeq
One can see that we have to suppose that the acceleration  rate is quite 
high $q\sim 0.1$ in order to continuously supply electrons which 
radiate in MeV band. The rough estimate (\ref{emax}) does not take into
account the geometry of electromagnetic field near the horizon. 
Since we are dealing with ordered electric and magnetic field configuration, 
the energies of electrons (and, correspondingly, the spectrum of synchrotron 
radiation) depend strongly on the mutual orientation of magnetic field 
$\vec B_\bullet$, electric field $\vec E_\bullet$ and particle velocity
$\vec v$. It is natural to expect that the angular distribution 
of the power and of the spectrum of the synchrotron radiation 
is highly anisotropic. 

Since the particles are accelerated along (or oppositely to) the  
direction of 
the external magnetic field and since the typical gamma-factors of electrons 
and 
positrons are $\Gamma_e> 10$
the synchrotron flux is concentrated within two oppositely directed cones 
with  opening angle 
$\theta< 1/\Gamma_e\sim 5^o$. Beaming within 
a narrow cone explains why the pair 
production in $\gamma\gamma$ interactions does not prevent the MeV 
$\gamma$-rays to escape from the emission region (see \cite{grb-review}).

Contrary to neutron stars, black holes can not support their own magnetosphere.
Indeed, free charges can not be emitted from the black hole surface, they can 
only be accreted from infinity or generated {\it in situ}. If the magnetic 
field near horizon is strong, the $\gamma$-quanta can supply 
free charges due to the pair production in external magnetic field. 
Propagation length of a $\gamma$-quantum of energy $\epsilon_\gamma$ 
in magnetic field $B_\bullet$ 
exponentially decreases with energy, 
$\lambda_{B\gamma}\approx 10^{-6}\lf[10^{12}\mbox{ G}/B_\bullet\ri] 
\exp\l(8m_e^3/(3B_\bullet\epsilon_\gamma)\r)$ cm,  
and reaches $\lambda_{B\gamma}\sim R_\bullet$ when $\epsilon_\gamma$ rises to 
\beq
\label{pp}
\epsilon_{\gamma\rightarrow e^+e^-}
\approx 0.1\frac{m_e^3}{B_\bullet}=2\times 10^6
\lf[\frac{10^{12}\mbox{ G}}{B_\bullet}\ri]\mbox{ eV}
\eeq
Thus, $\gamma$-quanta which form a high-energy tail of the synchrotron
spectrum $\epsilon_{\gamma}>\epsilon_{\gamma\rightarrow e^+e^-}$   
can not leave the emission region, they are 
immediately converted into $e^+e^-$ pairs. In fact, if the direction of 
propagation of photons is inclined at angle $\theta_\gamma$  to the 
direction of magnetic field,   the threshold of pair 
production can be higher by a factor of $1/\theta_\gamma$, as compared to 
(\ref{pp}). The spectrum of $\gamma$-rays which escape from the acceleration 
region depends strongly on the orientation of an observer relative to the 
direction of magnetic field: harder spectra are observed at smaller viewing
angles.

The pair production increases the density of $e^+e^-$ plasma near the
horizon. At the same time, absorption of particles by the horizon 
leads to the decrease of the plasma density. The 
density of charge right near the horizon is determined by the 
competition between pair production and accretion rates.
If the plasma density  in the acceleration 
volume exceeds certain critical density (see \cite{blandford})
\beq
\label{crit}
n_{cr}\sim \frac{B_\bullet}{eR_\bullet}= 10^{15} 
\lf[\frac{B_\bullet}{10^{12}\mbox{ G}}\ri]
\lf[\frac{10M_\odot}{M_\bullet}\ri]\mbox{ cm}^{-3}
\eeq 
redistribution of charges in the 
vicinity of horizon can lead to the neutralization
of the strong parallel electric field and decrease in the efficiency $q$ of 
the acceleration process. Minimal 
time scale at which the charge redistribution can happen is about the light
crossing time of the acceleration region. This scale defines the minimal 
variability time  of the system
\beq
T_{var}\ge \frac{R_\bullet}{c}=10^{-4}\lf[\frac{ M_\bullet}{10M_\odot}\ri]
\mbox{ s.} 
\eeq
If the back reaction of freshly created $e^+e^-$ pairs on the external 
electromagnetic field reduces the efficiency of acceleration $q$, the typical 
energy of $\gamma$ quanta decreases as $\epsilon_\gamma\sim q$. The decrease
in energy of synchrotron quanta leads to the decrease in the pair production 
rate since most of the quanta have energies below the pair production threshold
(\ref{pp}). In the absence of the pair production the density of $e^+e^-$
plasma will decrease rapidly because all the plasma falls behind 
the horizon. But as soon as the plasma density decreases, the strong 
electric field induced by the black hole rotation can again accelerate 
particles more efficiently. If the acceleration efficiency rises back, 
the energies of electrons and positrons increase 
and  synchrotron $\gamma$-quanta can again produce 
pairs in magnetic field. Thus, there is a ``critical'' regime of 
operation of the central engine in which the plasma density is close to 
the value given by Eq. (\ref{crit}). 

The particle density (\ref{crit}) is significantly lower than the density 
in the central engine estimated in the fireball model. In the 
fireball model one assumes that the total energy of the GRB $E_{tot}\sim 
10^{52}$ erg 
is initially 
deposited into matter particles 
in the volume of the size defined by the 
variability time scale $R_{var}\sim 10^7$ cm. In this case one obtains 
the particle density $n_{fb}\sim 
E_{tot}/(R_{var}^3\Gamma_{fb}m)\sim 10^{31\div 34}$ cm$^{-3}\gg n_{cr}$ 
($\Gamma_{fb}\sim 10^{2\div 3}$ is the 
typical gamma-factor of the particles and $m$ is the mass of the particles, 
which is equal to electron mass in the case of a fireball without barions 
\cite{piran} and 
to the proton mass in the case of a ``cannon ball'' \cite{cb}). 
At such high particle 
density the fireball is not transparent to the $\gamma$ radiation. The optical
depth is $\tau=\sigma_Tn_{fb}R_{var}=\sigma_TE_{tot}/(R_{var}^2\Gamma_{fb} 
m_e)\sim 10^{15\div 18}$ ($\sigma_T$ is the Thomson cross section). 
That is why one has to 
suppose that the observed $\gamma$ emission originates from the distances much 
larger than the size of the central engine: the optical depth decreases as
$\tau\sim R^{-2}$ and reaches $\tau\sim 1$ when the size of the fireball 
(or cannonball) 
becomes $R_{fin}\sim 10^{14\div 16}$ cm. The  mechanism under 
consideration differs from the fireball model: work done by rotation-induced 
electric field on a charged particle is not used to increase the particle 
energy. Instead, it is immediately radiated in the form of $\gamma$ quanta.  
That's why one does not need to assume large 
particle density in the central engine.

When the central engine operates in the ``critical'' regime (\ref{crit}) 
 luminosity of the black hole can be estimated as follows. Since the 
acceleration rate balances the energy loss rate, each electron 
radiates with power $P_e\sim qeB_\bullet=1.4\times 10^{13}q(B_\bullet/10^{12}$ 
G$)$ erg/s. The total
number of electrons in the acceleration volume is $N=(4\pi/3) R_\bullet^3
n_{cr}$. Thus, the total luminosity is
\beq
\label{Lcrit}
L_{cr}=\frac{4\pi}{3}qR_\bullet^2B_\bullet^2=1\times 10^{48}q
\lf[\frac{B_\bullet}{10^{12}\mbox{ G}}\ri]^2
\lf[\frac{M_\bullet}{10M_\odot}\ri]^2\mbox{ erg/s.} 
\eeq
 If the direction toward an
observer lies  inside the emission cone with opening angle $\theta=1/\Gamma_e
\sim 0.1$,
 the inferred equivalent spherical luminosity
is $L_{sph}=\Gamma_e^2L_{cr}\sim 10^{50}$ erg/s.   

Compactness of the emission region in the model under consideration rises
the question of possible self-absorption of the radiation. The 
self-absorption length, which is the inverse of the absorption coefficient
is given by \cite{ginzburg}
\beq 
\lambda_{sa}\sim 0.1 \frac{m_e^4\epsilon_\gamma^3}{e^4B_\bullet^2E_en_{cr}}
\eeq
Substituting $n_{cr}$ (\ref{crit}) and 
$E_e$ (\ref{ee})  into this equation and taking $\lambda_{sa}$ to be about 
the size of the emission region $\sim R_\bullet$ we
find that the self-absorption becomes 
important for photons with energies $\epsilon_\gamma\le 
\epsilon_{sa}$ with
\beq
\label{sa}
\epsilon_{\rm sa}\sim 30\left[\frac{B_\bullet}{10^{12}\mbox{G}}
\right]^{1/5} 
\mbox{ keV.}  
\eeq
Self-absorption of the low-energy part of $\gamma$ ray 
spectrum can be responsible for the very hard low-energy 
spectral index  $\alpha<-2/3$ of about a quarter of BATSE GRBs 
\cite{preece}. The possibility that the hard low-energy 
spectra can be due to 
the self-absorption in the source was considered in 
\cite{petrosian}. However, within the fireball-type models 
one has to assume unrealistic physical conditions in the 
shock \cite{granotpiran} if one tries to find the self-absorption 
for photons in the energy range (\ref{sa}).
  
{\it GRB afterglow and connection with supernovae explosions.}
The fate of the $\gamma$-ray beam produced in the central engine depends on 
the background through which it propagates toward an observer. The background 
in the vicinity of a newly born black hole is determined by the previous 
evolution of a progenitor star. Spectral characteristics of 
Type Ib/c supernovae (which are thought to be associated with GRBs) 
imply that 
progenitor stars experience an extended period of mass loss before 
the explosion
and completely loose their hydrogen envelope \cite{SNIc}. 
The ``bare core'' progenitor star of Type Ib/c supernova 
should be surrounded by the extended region containing  matter ejected in the
stellar wind. 

At the distance scale of the order of size of the core, 
$R\sim 10^{10}$ cm, 
the $\gamma$-quanta emitted from the black hole can interact with the rests 
of the progenitor star
while at larger scales $R\sim 10^{15}$ cm the $\gamma$-ray 
beam power can be absorbed in 
interactions with the wind particles.  
The geometry of the inflowing 
material  is expected to be highly complicated. Rotation of the
progenitor leads to faster infall of the polar regions as compared 
to the equatorial regions \cite{woosley}. Therefore
 we have to consider two possibilities: when the 
$\gamma$-ray beam can escape from the collapsed core and when the power of
the beam is absorbed in the rests of the core. Of course, it can happen that
just a fraction of the $\gamma$-ray power is absorbed in the core. 

If the $\gamma$-ray
beam interacts mostly with the extended wind region around the progenitor 
star, the propagation distance of the $\gamma$-quanta can be estimated as 
follows. Assuming the asymptotic velocity of the wind 
$v_\infty\sim 10^3$ km/s 
and the mass loss rate  $\dot M_{wind}\sim 10^{-5}M_\odot/$yr, 
we can estimate the density of background particles at the distance 
$D$ from the progenitor as
\beq
\label{nwind}
n_{wind}\approx 3\times 10^{11}\lf[\frac{
\dot M_\star}{10^{-5}M_\odot/\mbox{yr}}\ri]
\lf[\frac{10^{12}\mbox{ cm}}{D}\ri]^2
\mbox{ cm}^{-3}\mbox{.} 
\eeq
The total cross 
section of interaction of $\gamma$ quanta with wind  electrons 
can be estimated as $\sigma\sim\sigma_T$. 
The mean free path $\lambda_{\gamma,wind}=(\sigma n_{wind})^{-1}$ 
of $\gamma$-rays is equal to
\beq
\lambda_{\gamma,wind}\sim 1.5\times 10^{13}\lf[\frac{10^{11}}{n_{D}}\ri] 
\mbox{ cm}
\eeq
This means that significant part of the $\gamma$-ray flux can be absorbed 
during propagation through the wind zone. 
In this case a patch of the wind which is situated on 
the way of the $\gamma$ beam will receive an energy 
$E_{tot}\sim L_{cr}T_{GRB}$ where $L_{cr}$ (\ref{Lcrit}) is the beam power 
and $T_{GRB}$ is the duration of the GRB. 
Assuming $T_{GRB}\sim 10^{1\div 2}$ 
s we find that the wind patch can acquire  total
energy $E_{tot}\sim 10^{49\div 50}$ erg. If we compare this 
energy to the rest energy of the wind particles contained within the volume 
of the size $D\sim 10^{13}$ cm, $E_{rest}=Mc^2\sim 
(4\pi/3)D^3n_{wind}m_p=10^{47}$
erg (for the value of $n_{wind}$ given in (\ref{nwind})), 
we find that the transmitted energy is enough to accelerate 
the patch of the wind to the speed comparable to the speed of light. 
In this case the situation is similar to the one assumed in the 
fireball model: large energy $E_{tot}\gg Mc^2$ 
deposited into kinetic energy of the 
matter will be subsequently radiated as the 
GRB afterglow. Different models of evolution of relativistic ejecta, 
like relativistic fireball models \cite{piran}, or cannon ball model 
\cite{cb} can provide satisfactory fits to the properties of GRB 
afterglows.

If one or both $\gamma$-ray beams hit the rests of the progenitor star, 
rather than escape into the wind zone,  the energy of the beam can be 
absorbed within the star. As a result a relativistic jet will form inside
the (rests of the) collapsing core. This jet can give rise to the jet-induced 
supernova explosion \cite{jetSN}. The jet mechanism of supernova explosions 
can be generally valid in the case of highly asymmetric explosions 
of very massive stars (see  \cite{hegeretal}). Note that if the initial 
magnetic field near the black hole horizon is higher than $10^{13}$ G, the
typical gamma-factor of electrons is $\Gamma_e<10$ (\ref{emax}). In this case
the synchrotron emission is almost isotropic. The power $L_{cr}\sim
10^{50}$ erg/s (\ref{Lcrit}) is emitted in the  form of  
$\gamma$-quanta with energies below the pair production threshold 
$\epsilon_{\gamma\rightarrow e^+e^-}\le 0.1$ MeV (\ref{pp}). This powerful 
$\gamma$-ray flux can be absorbed in the outer parts of 
the collapsing core. Absorption of this 
isotropic flux can lead to the bounce and subsequent expansion of the accreting
material. In this case the behavior of ejected matter should be 
similar to the one considered in the ``dirty fireball'' or ``hypernova'' 
model discussed in Ref. \cite{paczynski}.

To summarize, we have considered a mechanism of the 
prompt $\gamma$-ray emission of
GRBs in which the beamed $\gamma$-ray flux is produced by 
electrons and positrons propagating in the strong magnetic field field 
$B_\bullet\sim 10^{12}$ G in the vicinity of a young black hole  
formed in the core collapse of a massive star. This mechanism can naturally
account for 
the high degree of polarization and hard low energy spectral index of
a part of GRBs. Since the 
$\gamma$-ray flux originates  directly in the central engine of GRB,  observed
characteristics of the GRBs are directly related to the 
physical properties of the central engine. The GRB photons are emitted 
from the vicinity of horizon where nonperturbative effects of General
Relativity are important. Besides, the synchrotron $\gamma$-quanta are 
emitted by electrons in magnetic field whose strength is just an 
order of magnitude below the critical value $B_{q}=4.4\times 10^{13}$ G 
at which quantum effects set on. Thus, within the considered model,
prompt emission of GRBs can provide an interesting laboratory 
for General Relativity and strong magnetic field physics.  
$\gamma$-quanta emitted form the central engine of GRB can be partially 
absorbed during the propagation through the extended wind 
zone around the progenitor star or through the  accreting/ejected matter.
Interaction of the $\gamma$-rays with background particles 
leads to production 
of relativistic ejecta (``fireballs'') 
from the central engine which are responsible 
for the GRB afterglows.

\end{document}